\documentstyle[prd,aps,eqsecnum]{revtex}
\begin{document}
\preprint{PU-RCG/98-5, gr-qc/9804046}
\draft
\renewcommand{\topfraction}{0.8}
\renewcommand{\bottomfraction}{0.8}
\twocolumn[\hsize\textwidth\columnwidth\hsize\csname 
@twocolumnfalse\endcsname
\title{Gauge-Invariant Variables on Cosmological Hypersurfaces} 
\author{Karim A. Malik and David Wands}
\address{School of Computer Science and Mathematics, University of 
Portsmouth, Portsmouth PO1 2EG,~~~U.~K.}  
\date{\today} 
\maketitle
\begin{abstract}
We show how gauge-invariant cosmological perturbations may be
constructed by an unambiguous choice of hypersurface-orthogonal
time-like vector field (i.e., time-slicing). This may be defined
either in terms of the metric quantities such as curvature or shear,
or using some matter field. As an example, we show how linear
perturbations in the covariant fluid-flow approach can then be
presented in an explicitly gauge-invariant form in the coordinate
based formalism.
\end{abstract}
\pacs{PACS numbers: 98.80.Hw \hfill Preprint  PU-RCG/98-5, gr-qc/9804046}
\vskip2pc]
 


In an unperturbed Friedmann-Robertson-Walker (FRW) universe the
homogeneous spatial hypersurfaces pick out a natural cosmic time
coordinate, and hence a (3+1) decomposition of spacetime. But in the
presence of inhomogeneities this choice of coordinates is no longer
unambiguous.

The need to clarify these ambiguities lead to the fluid-flow approach
which uses the velocity field of the matter to define perturbed
quantities orthogonal to the fluid-flow~\cite{Hawking,Lyth,EB}.
An alternative school has sought to define gauge-invariant
perturbations in any coordinate system by constructing quantities that
are explicitly invariant under general coordinate
transformations~\cite{Bardeen,KS,MFB}.
Results obtained in the two formalisms can be difficult to
compare. One stresses the virtue of invariance of metric perturbations
under gauge transformations, while the other claims to be
covariant and therefore manifestly gauge-invariant due to its
physically transparent definition.

The purpose of this brief report is to stress that linear
perturbations defined on an unambiguous physical choice of
hypersurface can always be written in a gauge-invariant manner. In the
coordinate based formalism~\cite{Bardeen} the choice of hypersurface
implies a particular choice of gauge, but by including the explicit
gauge transformation from an arbitrary initial coordinate system the
metric perturbations can be given in an explicity gauge-invariant
form.
Similar conclusions were presented in a recent paper by
Unruh~\cite{Unruh} while the present paper was in preparation.
In this language, linear perturbations in the fluid-flow or
``covariant'' approach appear as a particular gauge choice (the
comoving orthogonal gauge~\cite{Bardeen,KS}) whose metric
perturbations can be given in an explicitly gauge-invariant form.  But
there are many other possible choices of hypersurface, including the
zero-shear orthogonal (or longitudinal)
gauge~\cite{Bardeen,KS,MFB,Bertschinger} in which gauge-invariant
quantities may be defined.


\section{The metric approach}

We will restrict our analysis to the case where the perturbations can
be constructed from scalar quantities defined on 3-D
hypersurfaces~\cite{Bardeen,Stewart}. 
The line element allowing arbitrary linear scalar perturbations of a FRW
background can be written
\begin{eqnarray}
\label{ds}
ds^2 &=& a^2(\eta) \left\{ -(1+2\phi) d\eta^2+2B_{|i}d\eta dx^i
 \right. \nonumber \\
&& \ \ \left. +
\left[(1-2\psi)\gamma_{ij}+2E_{|ij} \right] dx^idx^j \right\} \, ,
\end{eqnarray}
where we use the conformal time $\eta$. The spatial metric on 3-spaces
of constant curvature is given by $\gamma_{ij}$ and covariant
derivatives with respect to this metric are denoted by $X_{|ij}$.
\footnote{Our notation coincides with that of Mukhanov, Feldman and
Brandenberger~\cite{MFB} which is widely used in
the literature. For comparison with the notation of
Bardeen~\cite{Bardeen} note that
\begin{eqnarray}
\phi \equiv AQ^{(0)} \, , 
&\qquad& 
\psi \equiv - \left( H_L+{1\over3}H_T \right) Q^{(0)} \ ,
\nonumber \\
B \equiv {B_B Q^{(0)} \over k} \ , 
&\qquad& 
E \equiv {H_T Q^{(0)} \over k^2} \, ,
\end{eqnarray}
where Bardeen explicitly included $Q^{(0)}(x^i)$, the eigenmodes of the
spatial Laplacian with eigenvalue $-k^2$.}

The intrinsic spatial curvature on hypersurfaces of
constant conformal time $\eta$ is given by \cite{Bardeen,KS}
\begin{equation}
^{(3)}R = {6\kappa\over a^2}
 + \frac{12\kappa}{a^2} \psi + \frac{4}{a^2} \psi_{|i}^{~i} 
\end{equation}
For a perturbation with comoving wavenumber $k$ we therefore have
\begin{equation}
\delta^{(3)}R = \frac{4}{a^2} \left(3\kappa-k^2\right)~\psi 
\end{equation}
and $\psi$ is often simply referred to as the curvature perturbation. 

The metric perturbations $\phi$, $\psi$, $E$ and $B$ can also be
related to various geometrical quantities defined in terms of
the unit time-like vector field
\begin{equation}
\label{Nmu}
N^\mu={1\over a}(1-\phi,B_{|}^{~i}) \,.
\end{equation}
which is orthogonal to the constant-$\eta$ hypersurfaces.
We can write the expansion, acceleration and shear of the vector
field~\cite{KS}
\begin{eqnarray}
\label{expansion}
\theta &=& 3{a'\over a^2} \left(1- \phi \right) - {3\over a} \psi' -
{1\over a} \left(B-E'\right)_{|i}^{~i} \,,\\
\label{acceleration}
a_i &=& \phi_{|i}\,, \\
\sigma_{ij} &=& a \left( \sigma_{|ij}
  - {1\over3} \gamma_{ij} \sigma_{|k}^{~k} \right) \,,
\end{eqnarray}
where the scalar describing the shear is
\begin{equation}
\label{shear}
\sigma = - B + E' \,.
\end{equation}
Note that all these physical quantities can be written in terms of just
three scalars $\phi$, $\psi$ and $\sigma$. 



The homogeneity of a FRW spacetime gives a natural choice of
coordinates in the absence of perturbations. But in the presence
of linear perturbations we are free to make a first-order change in
the coordinates, i.e., a gauge transformation,
\begin{equation}
\label{gauge}
\tilde\eta = \eta+\xi^0 \,, \qquad
\tilde x^i = x^i+\xi_{|}^{~i} \,,
\end{equation}
where $\xi$ and $\xi^0$ are arbitrary scalar functions.
A scalar transformation of this form preserves
the scalar nature of the metric perturbations~\cite{Bardeen,Stewart}
The function $\xi^0$ determines the choice of constant-$\eta$
hypersurfaces, i.e., the time-slicing, while $\xi$ then selects the
spatial coordinates within these hypersurfaces.  The choice of
coordinates is arbitrary to first-order and the definitions
of the first-order metric and matter perturbations are thus
gauge-dependent.

The coordinate transformation of Eq.~(\ref{gauge}) induces a change in
the functions $\phi$, $\psi$, $B$ and $E$ defined by Eq.~(\ref{ds})
\begin{eqnarray}
\label{transphi}
\tilde\phi&=&\phi-h \xi^0-\xi^{0\prime} \\
\label{transpsi}
\tilde\psi&=&\psi+h\xi^0 \\
\label{transB}
\tilde B&=&B+\xi^0 -\xi' \\
\label{transE}
\tilde E&=&E-\xi
\end{eqnarray}
where $h=a'/a$ and a dash indicates differentiation with
respect to conformal time $\eta$.
Any scalar $\varphi$ (including the fluid density or
pressure) which is homogeneous in the background FRW model can be
written as $\varphi(\eta, x^i) = \varphi_0(\eta) + \delta\varphi(\eta,
x^i)$.  The perturbation then transforms as
\begin{equation}
\label{transvarphi}
\delta \tilde\varphi = \delta \varphi - \xi^0 \varphi'_0 \, ,
\end{equation}
Physical scalars on the hypersurfaces, such as the curvature,
acceleration, shear or $\delta\varphi$, only depend on the choice of
$\xi^0$, but are independent of the coordinates within the 3-D
hypersurfaces determined by $\xi$. The function $\xi$ can only affect
the components of 3-vectors or 3-tensors on the hypersurfaces and not
3-scalars.



The gauge-dependence of the metric perturbations lead Bardeen to
propose that only quantities that are explicitly gauge-invariant under
gauge transformations should be considered. The two scalar gauge
functions allow two of the metric perturbations to be eliminated
implying that one should seek two remaining gauge-invariant
combinations. By studying the transformation
Eqs.~(\ref{transphi}--\ref{transE}), Bardeen constructed two
such quantities~\cite{Bardeen,MFB}\footnote
{In Bardeen's notation these gauge-invariant perturbations are given
as $\Phi\equiv\Phi_A Q^{(0)}$ and $\Psi\equiv-\Phi_H Q^{(0)}$.}
\begin{eqnarray}
\label{Phi}
\Phi &\equiv& \phi + h(B-E') + (B-E')'  \,,\\
\label{Psi}
\Psi &\equiv& \psi - h \left( B-E' \right) \,.
\end{eqnarray}
These turn out to coincide with the metric perturbations in a
particular gauge, called variously the orthogonal
zero-shear~\cite{Bardeen,KS}, conformal Newtonian~\cite{Bertschinger}
or longitudinal gauge~\cite{MFB}. 
It may therefore appear that this gauge is somehow preferred over
other choices. However any unambiguous choice of time-slicing can be
used to define explicitly gauge-invariant perturbations. The
longitudinal gauge of Ref.~\cite{MFB} provides but one example.


If we choose to work on spatial hypersurfaces with vanishing shear
$\tilde\sigma$ this implies that starting from arbitrary coordinates we
should perform a gauge-transformation
\begin{equation}
\xi^0_l = -B+E' \,.
\end{equation}
This is sufficient to determine the $\phi$, $\psi$, $\sigma$ or any
other scalar quantity on these hypersurfaces. 
In addition, the longitudinal gauge is completely determined by the
spatial gauge choice 
\begin{equation}
\xi_l = E \,,
\end{equation}
and hence $\tilde{E}=\tilde{B}=0$. 
The remaining functions $\phi$, $\psi$ and $\delta\varphi$ become
\begin{eqnarray}
\tilde \phi_l &=& \phi + h(B-E') + (B-E')' \,, \\
\tilde \psi_l &=& \psi - h \left( B-E' \right) \,, \\
%
%
\delta \tilde \varphi_l &=& \delta \varphi + \varphi_0'
\left(B-E'\right) \,.
\end{eqnarray}
Note, that $\tilde\phi_l$ and $\tilde\psi_l$ are then identical to
$\Phi$ and $\Psi$ defined in Eqs.~(\ref{Phi}) and~(\ref{Psi}).
These gauge-invariant quantities are simply a coordinate independent
definition of the perturbations in the longitudinal gauge.
{\em Other specific gauge choices may equally be used
to construct quantities that are manifestly gauge-invariant}.


An interesting alternative gauge choice, defined purely by local
metric quantities is the uniform curvature gauge~\cite{KS,Hwang,LS}, also
called the off-diagonal gauge~\cite{pbb}. In this gauge one selects
spatial hypersurfaces on which the induced 3-metric is left
unperturbed, which requires $\tilde\psi=\tilde{E}=0$. This corresponds
to a gauge transformation
\begin{equation}
\xi^0_\kappa = -{\psi\over h} \,, \qquad
\xi_\kappa = E \,.
\end{equation}
The gauge-invariant definitions of the remaining metric degrees of
freedom are then from Eqs.~(\ref{transphi}) and~(\ref{transB})
\begin{eqnarray}
\tilde\phi_\kappa &=& \phi + \psi + \left( {\psi\over h}
\right)^{\prime} \,, \\
\tilde{B}_\kappa &=& B-E'-{\psi\over h} \,, 
\end{eqnarray}
These gauge-invariant combinations were denoted ${\cal A}$ and ${\cal
B}$ by Kodama and Sasaki~\cite{KS}.
Perturbations of scalar quantities in this gauge become
\begin{equation}
\label{Mphi}
\delta\tilde\varphi_\kappa = \delta\varphi + \varphi_0' {\psi \over h}
\,.
\end{equation}
In some circumstances it is actually more convenient to use these
alternative gauge-invariant variables instead of $\Phi$ and
$\Psi$. For instance, when calculating the evolution of perturbations
during a collapsing ``pre Big Bang'' era
 the perturbations $\tilde\phi_\kappa$ and
$\tilde{B}_\kappa$ may remain small even when $\Phi$ and $\Psi$ become
large~\cite{pbb}. 
Note that Eq.~(\ref{Mphi}) gives the gauge-invariant scalar field
perturbation introduced by Mukhanov~\cite{Mukhanov}. 

For comparison note that the synchronous gauge, defined by
$\tilde\phi=\tilde{B}=0$, does not determine the time-slicing
unambiguously. There is a residual gauge freedom $\hat\xi^0=X/a$,
where $X(x^i)$ is an arbitrary function of the spatial coordinates,
and it is not possible to define gauge-invariant quantities in general
using this gauge condition~\cite{MS}.


\section{The fluid-flow approach}

Thus far we have concerned ourselves solely with the metric and its
representation under different choices of coordinates. However in any
non-vacuum spacetime we will also have matter fields to consider. Like
the metric, the coordinate representation of these fields will also be
gauge-dependent.

The stress-energy tensor of a perfect fluid with density $\epsilon$,
isotropic pressure $p$ and 4-velocity $u^{\mu}$ is given by
\begin{equation}
T^{\mu}_{~\nu} = \left( \epsilon +p\right) u^{\mu}u_{\nu}+p
\delta^{\mu}_{~\nu} .
\end{equation}
The linearly perturbed velocity can be written as
\begin{equation}
u^{\mu}=\frac{1}{a} \left( \left(1-\phi\right), ~v_{|}^{~i} \right),
\end{equation}
where we enforce the constraint
%
$u_{\mu}u^{\mu}=-1$. 
%
We can introduce the velocity potential $v$ since the flow
is irrotational for scalar perturbations.
We then get for the components of the stress energy tensor
\begin{eqnarray}
T^0_{~0} &=& -(\epsilon_0+\delta\epsilon)\\
T^0_{~i} &=&(\epsilon_0+p_0)~ \left(B +v \right) _{|i} \\
%
%
T^i_{~j} &=& (p_0+\delta p) ~\delta^i_{~j}+ \pi^i_{~j}
\end{eqnarray}
where we have included the trace-free anisotropic
stress tensor, $\pi^i_{~j}=\pi_{|~j}^{~i}- \frac{1}{3}
\delta^{i}_{~j}\pi_{|~k}^{~k}$.
%

Coordinate transformations affect the split between spatial and
temporal components of the matter fields and so quantities like the
density, pressure and 3-velocity are gauge-dependent. Density and
pressure are scalar quantities which transform as given in
Eq.~(\ref{transvarphi}), but the velocity potential becomes
\begin{equation}
\label{transv}
\tilde v = v + \xi' \,.
\end{equation}
The anisotropic pressure, $\pi_{ij}$, is gauge-invariant.


The comoving gauge is defined by choosing spatial coordinates such
that the 3-velocity of the fluid vanishes,
$\tilde{v}=0$. Orthogonality of the constant-$\eta$ hypersurfaces to
the 4-velocity, $u^\mu$, then requires $\tilde{v}+\tilde{B}=0$. From
Eqs.~(\ref{transB}) and~(\ref{transv}) this implies
\begin{eqnarray}
\xi^0_m &=& -(v+B) \nonumber\\
\xi_m&=& -\int v d\eta + \hat\xi(x^i)
\end{eqnarray}
where $\hat\xi(x^i)$ represents a residual gauge freedom,
corresponding to a constant shift of the spatial coordinates.
All the physical quantities like curvature, expansion, acceleration
and shear are independent of $\hat\xi$.
Applying the above transformation from arbitrary coordinates, the scalar
perturbations in the comoving orthogonal gauge can be written as
\begin{eqnarray}
\tilde \phi_m &=& \phi+\frac{1}{a} \left[ \left( v+B \right) a \right]'  \\
\label{psim}
\tilde \psi_m &=& \psi - h \left( v+B \right) \\
\tilde E_m &=& E + \int v d \eta - \hat\xi \\
\label{varphim}
\delta \tilde \varphi_m &=& \delta \varphi - \varphi'_0 \left(v+B\right)
\end{eqnarray}
Defined in this way, these combinations are gauge-invariant under
transformations of their component parts in exactly the same way as,
for instance, $\Phi$ and $\Psi$ defined in Eqs.~(\ref{Phi})
and~(\ref{Psi}), apart from the residual dependence of $\tilde{E}_m$
upon the choice of $\hat\xi$.

The density perturbation on the comoving orthogonal hypersurfaces is
given by Eq.~(\ref{varphim}) in gauge-invariant form as
\begin{equation}
\delta\tilde\epsilon_m = \delta\epsilon - \epsilon_0'
\left(v+B\right) \,,
\end{equation}
and corresponds to the gauge-invariant density perturbation
$\epsilon_mE_0Q^{(0)}$ in the notation of Bardeen~\cite{Bardeen}.  The
gauge-invariant scalar density perturbation $\Delta$ introduced in
Ref.~\cite{EB} corresponds to
$\delta\tilde\epsilon_{m|i}^{~~~i}/\epsilon_0$.

If we wish to write these quantities in terms of the metric
perturbations rather than the velocity potential then we
can use the Einstein equations~\cite{MFB} to obtain
\begin{equation}
v+B = \frac{h \phi + \psi' - \kappa \left(B-E'\right)}{h' - h^2 -
\kappa}
\,.
\end{equation}
In particular we note that we can write the comoving curvature
perturbation, given in Eq.~(\ref{psim}), in terms of the longitudinal
gauge-invariant quantities as
\begin{equation}
\tilde \psi_m
 = \Psi - \frac{h(h\Phi+\Psi')}{h'-h^2-\kappa} \,.
\end{equation}


Alternatively we could use the matter content to pick out uniform
density hypersurfaces on which to define perturbed quantities. Using
Eq.~(\ref{transvarphi}) we see that this implies a gauge transformation
\begin{equation}
\xi_\epsilon^0 = {\delta\epsilon \over \epsilon_0'} \,,
\end{equation}
and on these hypersurfaces the gauge-invariant curvature perturbation
is~\cite{Deru,MS}
\begin{equation}
\tilde\psi_\epsilon = \psi + h {\delta\epsilon \over \epsilon_0'}
\end{equation}
This coincides with $\zeta_{\scriptscriptstyle{\rm BST}}$ 
defined in Refs.~\cite{BST,MS}.

In Ref.~\cite{MFB} the gauge-invariant variable
$\zeta_{\scriptscriptstyle {\rm MFB}}$ is
defined as
\begin{equation}
\zeta_{\scriptscriptstyle {\rm MFB}}
 = \Phi + \frac{2}{3} \frac{ \Phi' + \Phi h}{(1+w)h},
\end{equation}
where $w\equiv p_0/\epsilon_0$.  On large scales (where we neglect
spatial derivatives) and in flat-space ($\kappa=0$) with vanishing
anisotropic stresses ($\pi^i_{~j}=0$, which requires that
$\Phi=\Psi$~\cite{MFB}) all three quantities $\tilde\psi_m$,
$\tilde\psi_\epsilon$ and $\zeta_{\scriptscriptstyle {\rm MFB}}$
coincide. The curvature perturbation, in one or other of these forms,
is often used to predict the amplitude of perturbations re-entering
the horizon scale during the radiation or matter dominated eras in
terms of perturbations that left the horizon during an inflationary
epoch, because they remain constant on super-horizon scales (whose
comoving wavenumber $k\ll h$) for adiabatic
perturbations~\cite{LL93}. However, only $\tilde\psi_\epsilon$ (or
$\zeta_{\scriptscriptstyle {\rm BST}}$) is constant on large scales,
in the presence of anisotropic stresses, or background spatial
curvature. Neglecting spatial gradients, it obeys the simple evolution
equation~\cite{GBW}
\begin{equation}
\tilde\psi_\epsilon'
 = \left( {p_0'\over\epsilon_0'} - {\delta p\over\delta\epsilon}
 \right) 3 h \tilde\psi_\epsilon \,.
\end{equation}
The pre-factor on the right-hand-side is gauge-invariant and vanishes
for adiabatic perturbations.

\section{Summary}

It is with some trepidation that we present yet another paper
attempting to clarify the gauge-dependence of cosmological
perturbations.
Nonetheless we feel that there is an important clarification
of the coordinate based approach that has been previously overlooked,
or at least left unstated.  If we use the value of any physical scalar
to unambiguously specify the gauge function $\xi^0$, and hence
the time-slicing of the perturbed spacetime, then we can write the
resulting scalar metric perturbations $\phi$, $\psi$, $\sigma$ or any
matter perturbation $\delta\varphi$ on this hypersurface in a
manifestly gauge-invariant way by explicitly including the
transformation from an arbitrary coordinate system. If in addition we
make an unambiguous choice of the spatial coordinates on these
hypersurfaces, through the gauge-function $\xi$, then all the 3-tensor
components also become gauge-invariant.

Examples of such gauge-invariant quantities can be constructed using
the zero-shear (longitudinal) or comoving orthogonal gauges.  One
advantage of the comoving or fluid-flow approach is that the (3+1)
decomposition need not be restricted to linear perturbations, and the
metric perturbations in the coordinate-basis appear as a linearisation
of the more general case~\cite{EB}. Realising that there are other
possible physical choices of hypersurfaces opens up the possibility of
considering non-linear perturbations in other coordinate systems, as
recently proposed by Sasaki and Tanaka~\cite{Sasaki}.


\end{document}